\documentclass[prb,aps,amssym,nofootinbib,floatfix,twocolumn,showpacs]{revtex4} 
\usepackage{epsfig,natbib}
\usepackage{graphicx}
\usepackage{amsfonts}
\usepackage{amsmath, amsthm, amssymb}
\usepackage{dsfont}
\usepackage{mathptm}
\usepackage{color}
\usepackage{hyperref}
\usepackage{tabularx}
\definecolor{Red}{rgb}{1,0,0}

\newcommand{\be}{\begin{equation}}
\newcommand{\ee}{\end{equation}}
\newcommand{\bea}{\begin{eqnarray}}
\newcommand{\eea}{\end{eqnarray}}

\begin{document}
\title{Spin dynamics of the anisotropic spin-1 antiferromagnetic chain at finite magnetic fields}
\author{Yousef Rahnavard}
\affiliation{Institut f\"ur Theoretische Physik, Technische Universit\"at Braunschweig, 38106 Braunschweig, Germany}
\author{Wolfram Brenig}
\affiliation{Institut f\"ur Theoretische Physik, Technische Universit\"at Braunschweig, 38106 Braunschweig, Germany}
\date{\today}
\begin{abstract}

We present results of a study of the antiferromagnetic spin-1 chain,
subject to the simultaneous presence of single-ion anisotropy and external
magnetic fields. Using quantum Monte-Carlo based on the stochastic series
expansion method we first uncover a rich quantum phase diagram comprising
N\'eel, Haldane, Luttinger liquid, and large anisotropy phases. Second, we
scan across this phase diagram over a wide range of parameters, evaluating
the transverse dynamic structure factor, which we show to exhibit sharp
massive modes, as well as multi particle continua. For vanishing anisotropy
and fields, comparison with existing results from other analytic and
numerical approaches shows convincing consistency.

\end{abstract} 
\maketitle

\section{Introduction}\label{sec:introduction}

Ever since Haldane's conjecture \cite{Haldane1983} on the difference between
even and odd half-integer Heisenberg antiferromagnetic spin chains (HAFC), the
spin-1 HAFC (S1-HAFC)
\be H=J\sum_n\vec S_n\cdot \vec S_{n+1}-h\sum_n S_n^z+D\sum_n (S_n^z)^2,
\label{wb1}
\ee
has been considered to be one of the fundamental models of low-dimensional
quantum magnetism. Here, the first term on the right hand side of (\ref{wb1})
refers to the {\em bare} chain, with antiferromagnetic exchange interaction $J$
and spin-1 operators $\vec S_n$ at sites $n$, and the remaining terms capture
common perturbations by single-ion anisotropy $D$ and external longitudinal
magnetic fields $h$.

For the isotropic case at zero magnetic field, i.e. $D= 0$ and $h= 0$, both,
static and dynamic properties of the S1-HAFC have been investigated extensively
using various theoretical and numerical methods \cite{Kolezhuk2004}. On the zone
boundary at $q=\pi$, its lowest-lying excitation is a massive 'single-magnon'
mode which displays the famous Haldane gap of $\Delta\simeq 0.41J$ \cite{Nightingale1986,White1993}
Near $q= 0$ the spectrum
comprises primarily a two-particle continuum of small spectral weight
\cite{Gomez-Santos1989,Affleck1992,Yamamoto1993,White1993}
This continuum is separated from the
ground state by $2 \Delta$. Finally, the next to dominant excitations near
$q=\pi$ are contained in a three-particle continuum starting at $3 \Delta$.
Theoretically these excitations have been obtained from several analytic
approaches, eg., mean-field theory \cite{Affleck1992},
the nonlinear $\sigma$ model
(NL$\sigma$M) \cite{Affleck1999}, as well as numerical methods, eg.,
quantum Monte-Carlo (QMC) \cite{Takahashi1989,Deisz1990,Meshkov1993,Yamamoto1997},
density matrix renormalization group (DMRG) \cite{White1993,White2008}.

Experimentally, the massive magnon at $q=\pi$ has been confirmed irrevocably,
however the two-, and in particular the three-particle continua remain a matter
of active research \cite{Ma1992,Zaliznyak2001,Kenzelmann2001}.

Most spin-1 chain compounds, such as NENP \cite{Renard1987,Regnault1994}, DTN \cite{Paduan-Filho2004,Zapf2006}, NENC
\cite{Orendac1995}, NDMAP \cite{Honda1998,Zheludev2001}, and NDMAZ \cite{Honda1997,Zheludev2001a} display sizable anisotropy $D$
and even the well studied prototype material $CsNiCl_3$ \cite{Steiner1987,Kenzelmann2002} has $D\neq 0$.
Therefore, it is of great interest to analyze the ground state properties and
the evolution of the excitation spectrum as a function of anisotropy. In
addition, many experimental studies, including neutron scattering
\cite{Facility1994,Zheludev2004} (NS), electron-spin resonance
\cite{Sieling2000,Zvyagin2008} (ESR), nuclear magnetic resonance
\cite{Chiba1991,Takigawa1996,Haga2000} (NMR), and thermal transport
\cite{Sologubenko2008} are performed at finite magnetic fields.

For vanishing magnetic field several studies have already been
performed regarding the quantum phases as a function of the anisotropy
\cite{Nijs1989,Schulz1986,Kennedy1992,Chen2003,DegliEspostiBoschi2003,Tzeng2008a,Hu2011,Furuya2011,Albuquerque2009}.
Similarly the magnetic field driven transition into a Luttinger liquid
(LLQ) phase \cite{Affleck1991,Golinelli1992,Konik2002} at $D=0$ is
well investigated. At finite $D$ {\em and} $h$ there are some studies
with planar or a combination of planar and axial magnetic fields and
additional other components of anisotropies
\cite{Pollmann2010,Gu2009,Sengupta2007}, still, too little is
known about the region of finite $D$ and  $h$ for the Hamiltonian
(\ref{wb1}). Therefore, the central goal of this paper is to shed more
light onto the combined impact and interplay between finite anisotropy
and magnetic fields in S1-HAFCs, both regarding static and dynamic
properties.

The paper is organized as follows. In section \ref{sec:PD} we uncover and
discuss the quantum phase diagram of (\ref{wb1}) over a substantial range of $D$
and $h$. In section \ref{sec:DSF} we detail our results for the transverse
dynamic structure factor (tDSF) and analyze its evolution in terms of anisotropy
and magnetic fields. Where available comparison to other methods, in particular
NL$\sigma$M model calculations \cite{Affleck1992,Affleck1999} and tDMRG
\cite{White2008} will be provided. We conclude and summarize our findings
in Sec.\ref{sec:con}. Appendix \ref{sec:method} contains a short summary of
the quantum Monte-Carlo method we use.

\section{Quantum Phase diagram}\label{sec:PD}

In this section, and before analyzing its dynamical properties, we will evaluate
the ground state phase diagram of the chain versus single-ion
anisotropy and magnetic field.

At zero magnetic field, the phase diagram in terms of anisotropy has been
already investigated\cite{Schulz1986,Nijs1989,Kennedy1992,Chen2003}.  It
consists of a N\'{e}el, a Haldane, and a large-D phase. The transition from the
N\'eel to the Haldane phase is of Ising type, while that from the Haldane to
large-D phase is of Gaussian type.  The critical values $D_c$ for the transition
between these phases have been determined using various numerical methods including
exact diagonalization  \cite{Chen2003}, DMRG \cite{DegliEspostiBoschi2003,Tzeng2008a,Hu2011}, 
series expansions  and QMC \cite{Albuquerque2009}.
Although there are slight quantitative differences between the results from
different methods for $D_c$, it is generally believed that the transition
between the N\'eel and Haldane phase occurs around $D_{NH}\simeq -0.31 J$ and that
between the Haldane and large-D phase around $D_{HL}\simeq 1.0 J$
\cite{DegliEspostiBoschi2003,Tzeng2008a,Albuquerque2009,Hu2011}.

All three phases, N\'{e}el, Haldane, and large-D, are gapful. It is
known that the spin gap of Haldane phase decreases upon increasing the
easy-plane anisotropy up to the critical point $D_c$ and increases
again afterwards, however, one remains in a gapful state
\cite{Hu2011}.  Applying an external magnetic field can also suppress
the spin gap of the chain.
\cite{Affleck1991,Golinelli1992,Konik2002}.  This means that at large
enough fields, the S1-HAFC enters a LLQ phase.  Now the question is
how and where exactly the gap closes. In other words the boundaries of
this LLQ phase are to be obtained.  Here we will quantify this and
establish how in general the quantum phase diagram evolves within the
$D, h$ plane.

There are several ways to determine the extent of the Haldane phase. One is
to evaluate the string order parameter \cite{Nijs1989,Kennedy1992}, which
is a nonlocal probe of the topological order. This order parameter is
fragile with respect to perturbations which break rotation symmetry, while
keeping other symmetries such as time-reversal, parity and translation
symmetries intact \cite{Anfuso2007,Gu2009,Pollmann2010}. Since both, the
Haldane and large-D phases are gaped while the LLQ formed between them is
gapless, another and rather direct way to determine the boundary between
these phases is to scan the energy gap versus $D$, fixing eg. $h$. To
obtain the gap, we first evaluate the uniform spin susceptibility in
terms of temperature $\chi(T)$. We then extract the gap by fitting the
low-temperature values of $\chi(T)$ to $\chi(T)\approx
e^{-\Delta/T}P_{k}^{l}(T)/T$, where $P_{k}^{l}(T)$ is a Pad\'{e}
approximant of order $[k,l]$. In principle, finite size scaling of the gap
determined this way should be performed, in particular because of critical
behavior near the transition points \cite{Barber1983}. In pratice however
and because of the additional approximation introduced by the Pad\'{e}
fitting we simple use a sufficiently large system of $512$ sites.  The
lowest temperature we have considered is $T= 0.0078 J$.

\begin{figure}[tb]
\begin{center}
\includegraphics*[width=7.0cm]{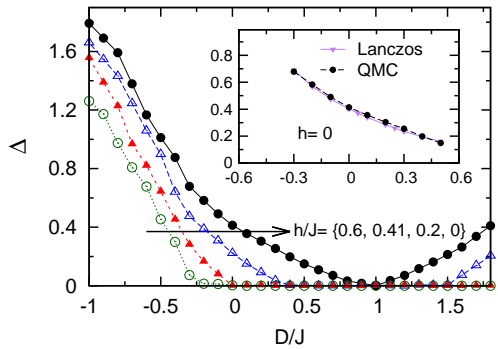}
\caption{Spin gap of the spin-1 chain in terms of single-ion anisotropy for
different external magnetic fields. Inset: for zero magnetic field, the
spin gap in terms of single-ion anisotropy obtained from QMC and Lanczos
method is shown.  Lanczos data are extracted from
Ref.\onlinecite{Furuya2011}. The system size and the lowest temperature considered for QMC data are $L=512$ and $T=0.0078$.}
\label{fig:spin-gap}
\end{center}
\end{figure}

\begin{figure}[tb]
\begin{center}
\includegraphics*[width=7.0cm]{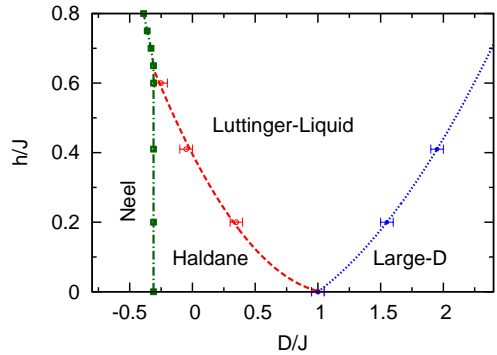}
\caption{Quantum phase diagram of the spin-1 Heisenberg chain versus
  single-ion anisotropy and magnetic field. The error bars of the
    transition points are equal to the distance between two sequential
    values of $D/J$ in each of the two procedures, gap or N\'{e}el
    order study.  In case of Haldane-N\'{e}el case, these error bars
    are smaller than the symbols.}
\label{fig:diagram}
\end{center}
\end{figure}

Fig. \ref{fig:spin-gap} summarizes all gaps $\Delta(D,h)$ extracted from
the preceding procedure, both, versus $D$ and for several magnetic
fields. For each gap, the Pad\'{e} fitting errors are found to be
within the QMC's error bars which are of the order of $10^{-4}$. Above the
critical field $h_c(D)$, there are two points of gap closure and reopening,
which we identify with the transition from the Haldane into the LLQ, and
from the LLQ into the large-D phase. At $D=D_{HL}$ the latter two points
have to merge at $h=0$ where the direct transition from the Haldane to
large-D occurs without accessing a LLQ phase. The gaps in
Fig. \ref{fig:spin-gap} display some clearly visible, albeit small,
noise. This is not related to QMC, or Pad\'{e} approximant errors. In fact,
for each individual gap extracted, the Pad\'{e} fitting errors are within
the QMC's error bars which are of the order of $10^{-4}$. Rather, the noise
is a consequence of the arbitrariness in choosing the upper cut-off for the
temperature range, used in Pad\'{e} fitting $\chi(T)$. This noise translates into an
error for the phase boundaries, which we find to dominate any corrections
from finite scaling. This justifies our neglect of the latter {\it a
posteriori}.

Previous studies \cite{Furuya2011} have analyzed the spin gap for $h=0$
using Lanczos spectra of small systems $L{\leq} 20$, over a restricted
range of $D_{NH}{<}D{<}D_{HL}$. As compared to QMC, finite size effects are
a relevant issue for this approach, and careful scaling analysis is
necessary, in particular for $D$ in the vicinity of $D_{HL}$, where gap
closure occurs. The inset of Fig. \ref{fig:spin-gap}, compares our
thermodynamic QMC gap with that obtained from extrapolating to
$L\rightarrow\infty$ in Ref. \onlinecite{Furuya2011}. The agreement is satisfying.

In Fig. \ref{fig:diagram} we collect the gap closure points obtained
from Fig. \ref{fig:spin-gap} as part of a quantum phase diagram versus
$D$ and $h$. The transition points are regarded as the midpoints of
the two sequential $D$ values for one of which $\Delta$ is finite
(gapped phase) while for the other one $\Delta \approx 0$ (gapless
phase).  Since the distance between two sequential $D/J$ values is
$0.1$, the uncertainty for the transition points is $\pm
0.05$.  The lines connecting the points are low order polynomials
fitted to the points. This figure also shows a transition from the
Ne\'{e}l to the Haldane phase, to which we turn now. Since both of the
latter phases are gapful, the transition line cannot be obtained from
a study of gap closures. Instead, we use that the long-range staggered
spin-correlation is an order parameter of the Ne\'{e}l phase, and
remains finite therein, while it decays in the Haldane phase
\cite{Charrier2010}. This has been applied previously to characterize
the Haldane phase as a function exchange anisotropy in the $XXZ$
spin-1 chain \cite{Su2012}.  The staggered spin-correlation reads
\begin{equation}
 {\cal O}^{\, z}_\textrm{N} (i,j)
 = (-1)^{i-j}\langle S^z_i S^z _j\rangle,
\end{equation}
and Ne\'{e}l ordering implies $O^{\, z} _\textrm{N} = \lim_{|i-j|\rightarrow
\infty}{\cal O}^{\, z}_\textrm{N} (i,j) \neq 0$.

\begin{figure}[tb]
\begin{center}
\includegraphics*[width=7.0cm]{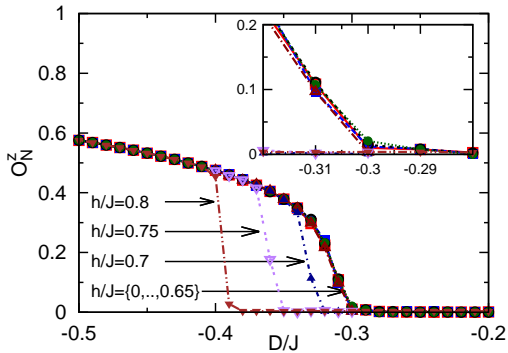}
\caption{N\'{e}el order parameter is shown as a function of anisotropy for
different magnetic fields. The inset shows the transition point between 
Haldane and Ne\'{e}l phase for $h/J={0,..,0.65}$ .
The system size and the temperature considered here are $L=512$ and $T= 0.01 J$.}
\label{fig:NeelCorr}
\end{center}
\end{figure}

Fig. \ref{fig:NeelCorr} displays our results for ${\cal O}^{\,
  z}_\textrm{ N} (0,L)$ for $L=512$. In view of the rather large
system, we refrain from finite size scaling analysis, and approximate
${\cal O}^{\, z}_\textrm{ N} (0,512)\simeq O^{\, z} _\textrm{N}$.  We
identify the phase transition point with the average value between the
smallest $D$ at which $O^{\, z} _\textrm{N}\approx 0$, where we are in
the Haldane phase, and the largest $D$ point at which $O^{\, z}
_\textrm{N} $ is finite, where we are in the Ne\'{e}l phase.  The
error in this case is $\pm 0.005$.  From this, and for
$h=0$ we obtain a transition point at $D_{NH}\approx -0.305 J$, which
is satisfyingly close to the values obtained by DMRG in
Refs. \onlinecite{Tzeng2008a,Hu2011}

The main message of Fig. \ref{fig:NeelCorr} is contained in the remarkable
evolution of $O^{\, z} _\textrm{N}$ and the quantum critical point with
magnetic field. As the figure shows, the critical value of $D_{NH}(h)$ is
independent of the field up to a critical value $\tilde{h/J}\approx 0.65$,
below which all curves for $O^{\, z} _\textrm{N}$ collapse onto a single
one. Adding $D_{NH}(h)$ into Fig. \ref{fig:diagram} shows that the point
($D_{NH}$,$\tilde{h}$) lies on the extrapolated line, approximating the
boundary of the LLQ phase. From the slope of this phase boundary, and since
$O^{\, z} _\textrm{N}$ has to be zero in the LLQ phase, further increasing
the field, the value of $D_{NH}(h)$ must decrease for $h{>}\tilde{h}$. This
is consistent with $O^{\, z}_\textrm{N}$ in Fig. \ref{fig:NeelCorr}. In
fact, as is obvious from the green squares in Fig. \ref{fig:diagram}, to
within the uncertainty of the LLQ phase boundary, the Ne\'{e}l-Haldane
transition is replaced by a direct transition from the Ne\'{e}l to the
Luttinger phase for $h>\tilde{h}$.

\section{Dynamic structure factor}\label{sec:DSF}

In this section we discuss the transverse dynamic structure factor
$\bf{S(q,\omega)}$. We will be interested in the Ne\'{e}l, Haldane, and
Luttinger phase. For this we analyze several values of magnetic field and
anisotropy, as shown in Fig. \ref{fig:diagram}.

\begin{figure}[tb]
\begin{center}
\includegraphics*[width=7.0cm]{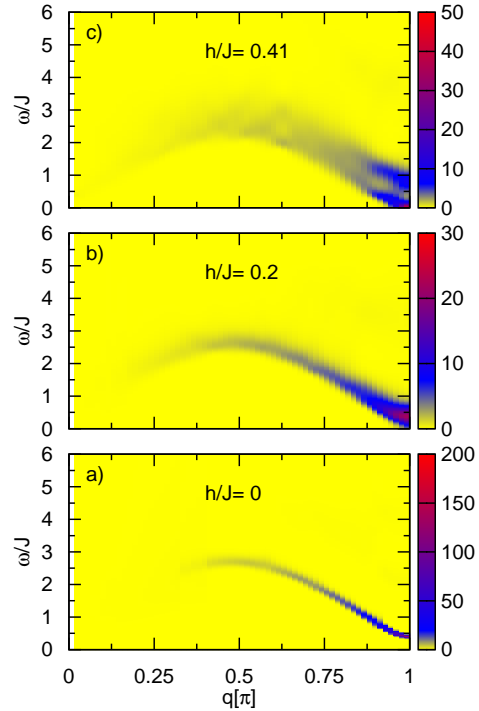}
\caption{Contour plot of the transverse dynamic structure factor of the
isotropic spin-1 chain as a function of frequency $\omega$ and wave vector
$q$ for three different magnetic fields $h=0, 0.2,$ and $0.41J$.  The
system size of the chain for all cases is $L=128$ and temperature is set to
$T= 0.1 J$. 
}
\label{fig:ContourDSFs}
\end{center}
\end{figure}

First, we focus on the field dependence of $\bf{S(q,\omega)}$ at the
isotropic point. Results for this are shown in the contour plots of
Fig. \ref{fig:ContourDSFs}. At zero magnetic field most of the spectral
weight is contained in a single, well defined excitation, which is clearly
visible in Fig. \ref{fig:ContourDSFs}a).  Most of the spectral weight of this so-called one
magnon mode resides at large momenta near $q=\pi$ and decreases rapidly
towards lower momenta, where we find that the integrated weight is
proportional to $q^2$ as $q\rightarrow 0$. 
At finite magnetic field, the two triplet branches which can be reached $\Delta S = +1$
transitions, split according to their Zeeman energy. For small fields, this
splitting is manifest through a broadening of the one-magnon line, while
the intensity decreases, as can been seen from Fig. \ref{fig:ContourDSFs}b) and by comparing
their intensity scales. If the Zeeman splitting is larger than the
broadening of the one-magnon excitations due to thermal, interaction, and
MaxEnt effects, then the splitting is directly visible, as in Fig. \ref{fig:ContourDSFs}c). In
that panel, the Zeeman energy has been chosen identical to the Haldane
gap. As can be seen, at this point the maximum intensity of the lower
branch extrapolates to zero energy at $q=\pi$, i.e. the gap closes,
consistent with entering the LLQ phase.

\begin{figure}[htbp]
\begin{center}
\includegraphics*[width=9.0cm]{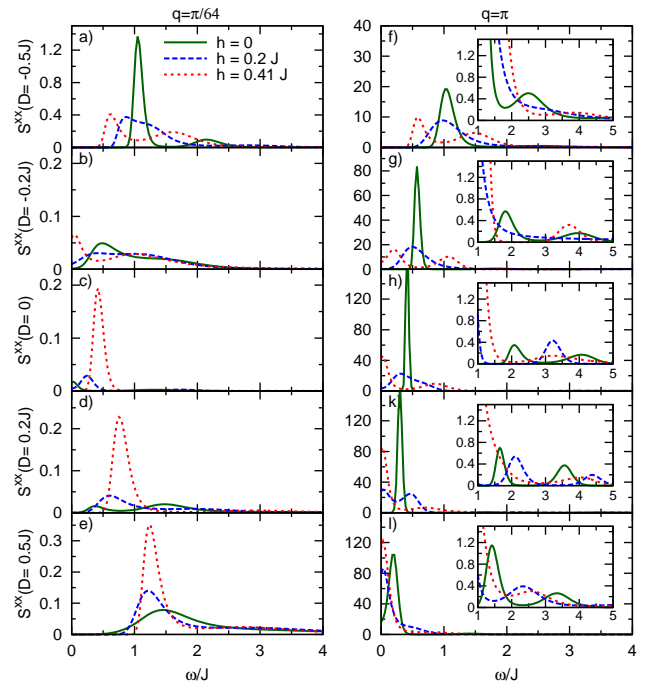}
\caption{Transverse dynamic structure factor of the spin-1 chain as a function of frequency at two 
q vectors for different anisotropies and magnetic fields is plotted.
The system size of the chain for all cases is $L=128$ and temperature is
set to $T= 0.1 J$.
}
\label{fig:DSFs}
\end{center}
\end{figure}

Next we focus on a more detailed discussion of 2D cuts of the spectra
versus $D$ and $h$ at small and large momenta. The main motivation for this
is, that apart from the one-magnon excitation, there are also
multi-particle continua present in the spectrum. While the former are most
dominant at large momenta, the latter exhibit very small spectral weight,
which remains invisible in contour plots of the full BZ, and can be
observed best at small momenta. Therefore we have plotted cuts through the
tDSFs at various anisotropies and magnetic fields in Fig. \ref{fig:DSFs},
which allows to clearly see the shape and weight of the spectra, despite
very large differences in their absolute scales in different frequency and
momentum regions. We have chosen three magnetic fields $h= 0$, $0.2$, $0.41
J$ and five different anisotropies, $D=-0.2, -0.5, 0, 0.2, 0.5$ ranging
from easy-axis to easy-plane anisotropies. For each of these cases two wave
vectors have been considered, i.e., $q=\pi$ (right panel), and $q= \pi/64$
(left panel), which is the smallest on the system for which we have
evaluated the tDSF, i.e. $L=128$.

We start with the isotropic case at small momenta,
Fig. \ref{fig:DSFs}c). At zero magnetic field it is dominated by a peak at
zero frequency. This {\it central peak} intensity stems from $\Delta S =+1$
transitions within thermally excited states and decreases as the
tempearture is lowered. In addition, there exists a continuum of very small
weight at higher frequencies, which however is not observable on the scale
of this plot. We emphasize, that this observation is rather distinct from
expectations \cite{White2008} at {\em zero} temperature, where the latter
multi-magnon spectrum should dominate the spectrum at low-momentum
displaying a gap of twice the Haldane gap. Increasing the wave vector, the
weight of the continuum gets larger as we will discuss later. Increasing
the magnetic field, the central peak and the continuum shift to larger
frequencies, with an energy scale set by the Zeeman energy.

Turning to finite anisotropy either of easy-axis type in
Fig. \ref{fig:DSFs}a) and b), or of easy-plane type in
Fig. \ref{fig:DSFs}d) and e), a shifting the dominant weight of
spectrum to larger frequencies is clearly visible. In addition to that an
interesting interplay between the effects of anisotropy and magnetic field
on the spectrum can be observed, which differs between the two kinds of
anisotropies. While in the case of easy-plane anisotropy, the magnetic
field only slightly shifts the weight of the spectrum, in the case of
easy-axis anisotropy, a splitting of the dominant peak result which
increases with increasing field.

Another interesting feature is the evolution of the spectrum versus
anisotropy upon switching from the Haldane into the Ne\'{e}l phase. In
the former we expect a rather broad continuum from multiparticle
excitations at small momentum, while in the latter a single dominant
excitation should occur, representing one-magnon excitations, which
may however still exhibit some broadening due to finite temperature
and interaction effects. Exactly this can be seen in going from
Fig. \ref{fig:DSFs}b) to a), where also the overall amplitude scale
increases by one order of magnitude in going from b) to a).  At
$q=\pi$, however, the spectrum gets broadened, as one goes from the
Haldane into the Ne\'{e}l phase by changing the anisotropy.

Moving to spectra at $q=\pi$, one can clearly see the sharp magnon peak
which dominates the spectrum at all anisotropies and magnetic fields.  At
zero magnetic field, the peak position which is a fingerprint of spin gap
shifts towards lower frequencies as we go from strong easy-axis anisotropy
$D= -0.5 J$ in Fig. \ref{fig:DSFs}f) to strong easy-plane one $D= 0.5 J$ in
Fig. \ref{fig:DSFs}l).  We find that in all cases studied there is good
agreement between the position of the peaks maximum and the thermodynamic
spin gaps which we have obtained in section \ref{sec:PD}. As can be seen
from \ref{fig:DSFs}f) through l) the monotonous behavior of the spectrum in terms
of anisotropy remains intact, even at finite magnetic field, but it is
accompanied by a splitting of the dominant peak due to Zeeman effect as we
have already mentioned in the context of Fig. \ref{fig:ContourDSFs}. The
splitting increases until the lower branch reaches its maximum at zero
frequency, at $h=h_c(D)$, where the LLQ is entered. Increasing the
magnetic field beyond $h_c(D)$ leads to an accumulation of spectral weight
at zero frequency and a depletion and smearing of the upper Zeeman
peak. This behavior is clearly visible in \ref{fig:DSFs} k) and l). In view of the phase
diagram Fig. \ref{fig:diagram}, and because the largest field we consider
in Fig. \ref{fig:DSFs} is $h_c(0)$, the maximum of the lower brance has to
stay above $\omega=0$ for $D<0$, which is exactly what we find in Fig. \ref{fig:DSFs}g)
and f).

In addition to the dominant single-magnon peak at $q=\pi$, we find a very
weak multi-particle continuum at higher frequencies. This is most likely
due to three-magnon excitations, as proposed in
Refs. \onlinecite{Affleck1999,White2008}. In view of the relative
intensities, it is remarkable, that our MaxEnt calculations are able to
resolve this continuum with respect to the single-magnon peak. Moreover
these results hint as to why experimental inelastic struture factor
determination of such continua have failed so far \cite{Zaliznyak2001}.

\begin{figure}[tb]
\begin{center}
\includegraphics*[width=7.0cm]{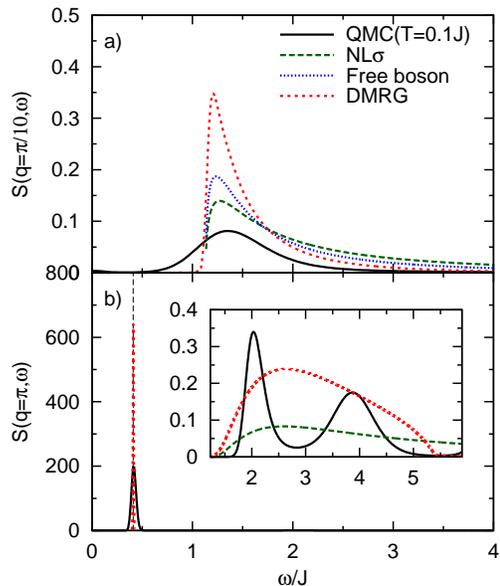}
\caption{ Dynamic structure factor of the isotropic chain at zero magnetic
field for two wave vectors ($q=\pi/10, \pi$) and different methods is
shown. For QMC data, the temperature is set to $T= 0.1 J$ while for other
data which are extracted from
Refs. \onlinecite{Affleck1992,Affleck1999,White2008} the temperature is set
to zero.}
\label{fig:comparison}
\end{center}
\end{figure}

Finally we contrast our findings with those from other approaches, namely
the NL$\sigma$ model and a free boson method
\cite{Affleck1992,Affleck1999}, as well as tDMRG \cite{White2008}. In
Fig. \ref{fig:comparison} results are shown for two momenta,
i.e. $q=\pi/10$ and $q=\pi$. For the former, the spectrum is dominated by
only the two-particle continuum and Fig. \ref{fig:comparison}a)
demonstrates good qualitative agreement between all different
methods. Regarding the quantitative difference of the QMC to the other
approaches, it is clear that the sharp onset of the continuum is
smeared. The reason for this is twofold, primarily it results from the fact
that our QMC is a finite temperature result, at $T=0.1 J$, while all other
methods refer to zero temperature. Additionally MaxEnt cannot be avoided to
introduce additional smoothing of any QMC spectrum. The spectrum at $q=\pi$
is shown in Fig. \ref{fig:comparison}b). As discussed previously, this
spectrum contains two largely separted intensity scales, one due to the
single-magnon mode, the other due to the three-particle continuum. For the
former and as for Fig. \ref{fig:comparison}a) we see convincing agreement between all
approaches for the locations of the magnon peak, including some finite
temperature and MaxEnt broadening of the QMC. Turning to the high energy
contiunuum at this wave vector, we first note that all methods result in a
comparable spectral intensity, however clear qualitative differences are
obvious. While both, NL$\sigma$ model and tDMRG display only a single
'hump', QMC results in two, moreover, while the tDMRG and QMC spectrum
remains confined to $1.5\lesssim\omega/J\lesssim 5.5$, the spectrum from
the NL$\sigma$ model continues up to much higher energies. The origin of
these differences remains unclear at present.

\section{Conclusions}\label{sec:con}

We have used QMC to study the antiferromagnetic spin-1 chain subject to the
simultaneous presence of single-ion anisotropy and external magentic
fields. The focus has been on two issues, namely the quantum phase diagram
and the transverse dynamic structure factor. We have uncovered a rich set
of quantum phases within the parameter range investigated, comprising
Ne\'{e}l, Haldane, Luttinger-liquid, and large anisotropy regimes. All
transitions where found to be of second order, as determined from either
closures of spin gaps, or by direct evaluation of order parameters.

Based on the phase diagram, we have determined the transverse spin
dynamics, covering the complete Brillouin zone, and varying the system
parameters to access the excitations within the Ne\'{e}l, Haldane,
Luttinger-liquid states. First, we have studied the spectral weight,
splitting and dispersion of the single-magnon mode, known from the standard
antiferromagnetic spin-1 chain, however versus anisotropy and external
fields. Second we have provided clear evidence for multi particle continua
with partially very small spectral weight and investigated their evolution
with momentum and system parameters.

Finally we have shown that our finite temperature spectra are consistent
with existing zero temperature results from other analytic as well as numerical
approaches. We hope that our findings may inspire additional experimental
studies using inelastic neutron scattering on spin-1 chain materials.

\acknowledgments

Part of this work has been supported by the Deutsche Forschungsgemeinschaft
through FOR912, Grant No. BR 1084/6-2, the European Commission through
MC-ITN LOTHERM, Grant No. PITN-GA-2009-238475, and the NTH School for
Contacts in Nanosystems. WB thanks the Platform for Superconductivity and
Magnetism Dresden, where part of this work has been performed, for its kind
hospitality.

\appendix
\section{Method}\label{sec:method}

All physical quantities in this paper are obtained using the stochastic
series expansion (SSE) method, pioneered by Sandvik $et\:al.$
\cite{Sandvik1992,Sandvik1999,Syljuasen2002}. This method is based on
importance sampling of the high temperature series expansion of the
partition function

\begin{equation}
Z=\sum_{\alpha}\sum_{n}\sum_{S_{n}}\frac{(-\beta)^{n}}{n!}\left\langle \alpha\right|\prod_{k=1}^{n}H_{a_{k},b_{k}}\left|\alpha\right\rangle \,\label{eq:partition}
\end{equation}
where $\beta$ is inverse temperature $1/T$ and
$H_{1,b}=1/2-S_{b1}^{z}S_{b2}^{z}$ and
$H_{2,b}=(S_{b1}^{+}S_{b2}^{-}+S_{b1}^{-}S_{b2}^{+})/2$ are spin diagonal
and off-diagonal bond
operators. $|\alpha\rangle=\left|S_{1}^{z},\ldots,S_{N}^{z}\right\rangle $
refers to the $S^{z}$ basis and
$S_{n}=[a_{1},b_{1}][a_{2},b_{2}]\ldots[a_{n},b_{n}]$ is an index for the
operator string $\prod_{k=1}^{n}H_{a_{k},b_{k}}$.  This string is
Metropolis sampled, using two types of update, diagonal updates
which change the number of diagonal operators $H_{1,b_{k}}$ in the operator
string and directed loop updates which change the type of operators
$H_{1,b_{k}}\leftrightarrow H_{2,b_{k}}$.
For nonfrustrated spin-systems the latter update comprises an
even number of off-diagonal operators $H_{2,b_{k}}$, ensuring positivity
of the transition probabilities.

The dynamic structure factor, is obtained from QMC in real
space $i,j$ and at imaginary time $\tau$. Following
Ref. \onlinecite{Sandvik1992} we consider
\begin{eqnarray}
S_{i,j}\left(\tau\right)=\Bigg\langle \sum_{p,m=0}^{n}\frac{\tau^{m}(\beta-\tau)^{n-m}n!}{\beta^{n}(n+1)(n-m)!m!}\times \phantom{aaa}\nonumber \\
S_{i}^{+}(p)S_{j}^{-}(p+m)\Bigg\rangle _{W}~,
\label{a2}
\end{eqnarray}
where $\langle\ldots\rangle_{W}$ refers to the Metropolis weight of an
operator string of length $n$ generated by the stochastic series expansion
of the partition function \cite{Sandvik1999,Syljuasen2002}, and $p,m$ are
positions in this string and $S_{i}(p)$ refers to the intermediate state
$\left|\alpha(p)\right\rangle=\prod^{p}_{k=1}H_{a_k,b_k}\left|\alpha\right\rangle$.

From (\ref{a2}) we proceed by performing a Fourier transformation into momentum space 
\begin{equation}
S({\bf q},\tau)=\sum_{a}e^{i{\bf q}\cdot{\bf r}_{a}}S_{a,0}(\tau)/L
\end{equation}
with $L$ being the system-size. The sought for form of the dynamical
structure factor in frequency and momentum space finally results from analytic
continuation to real frequencies based on the inversion of
\begin{equation}
 S({\bf q},\tau)=\frac{1}{\pi}\int_{0}^{\infty}d\omega S_{\perp}({\bf q},\omega)K(\omega,\tau),
\end{equation}
with a kernel $K(\omega,\tau)=e^{-\tau\omega}+e^{-(\beta-\tau)\omega}$.

The preceding inversion is an ill-posed problem, for which maximum entropy
methods (MaxEnt) have proven to be well suited. We have applied Bryan's
algorithm for our MaxEnt \cite{J.Skilling1984,Jarrell1996}. In a nutshell
this method minimizes the functional $Q=\chi^{2}/2-\alpha\sigma$, with
$\chi$ being the covariance of the QMC data with respect to the MaxEnt
trial spectrum $S({\bf q},\omega)$. Overfitting is prevented by an entropy
term $\sigma=\sum_{\omega}S({\bf q},\omega)\ln[S({\bf
q},\omega)/m(\omega)]$.  We have used a flat default model $m(\omega)$
which is iteratively adjusted to match the zeroth moment of the trial
spectrum. The optimal spectrum follows from the weighted average of $S({\bf
q},\omega)$ with the probability distribution $P[\alpha|S({\bf q},\tau)]$
adopted from Ref. \cite{J.Skilling1984}. Using static structure factors
evaluated by independent QMC runs, we have checked that all spectra
obtained from our MaxEnt, perfectly fulfill sum rules.

%\bibliographystyle{prsty}
%\bibliography{library}

\end{document}